\makeatletter
\@ifundefined{@parse@version@dash}{%
\def\@parse@version#1{\@parse@version@0#1}
\def\@parse@version@#1/#2/#3#4#5\@nil{%
\@parse@version@dash#1-#2-#3#4\@nil}
\def\@parse@version@dash#1-#2-#3#4#5\@nil{%
  \if\relax#2\relax\else#1\fi#2#3#4 }
}{}
\makeatother

\documentclass[preprintnumbers,aps,prd,twocolumn,superscriptaddress,%
   nofootinbib]{revtex4-2}
\usepackage{acronym}

\usepackage{epsf,amsmath,amssymb,graphicx,scalefnt,rotating,enumitem,fancyhdr}
\usepackage{dsfont}
\usepackage[english]{babel}
\usepackage{multirow}
\usepackage[final]{showkeys}
\usepackage{slashed}
\usepackage{filemod}
\usepackage[pdftex, colorlinks=true, linkcolor=black, filecolor=black,
  citecolor=black, urlcolor=black, draft=false, bookmarks,
  bookmarksnumbered=true, plainpages=false,
  linktocpage={true}]{hyperref}
\usepackage[usenames,dvipsnames]{xcolor}

\newcounter{notecount}

\newcommand{\lmut}{L_{\mu t}}

\newcommand{\Zchiring}{\mathring{Z}_\chi}

\newcommand{\ctr}{T_\text{R}}
\newcommand{\tr}{\ctr}
\newcommand{\nc}{n_\text{c}}

\newcommand{\nf}{n_\text{f}}

\newcommand{\ren}{\text{\abbrev{R}}}
\newcommand{\bare}{\text{\abbrev{B}}}
\newcommand{\citere}[1]{Ref.~\cite{#1}}
\newcommand{\citeres}[1]{Refs.~\cite{#1}}

\newcommand{\abbrev}[1]{{\scalefont{.9}#1}}
\newcommand{\EulerGamma}{\gamma_\text{E}}
\newcommand{\ep}{\epsilon}
\newcommand{\api}{a_s}

\newcommand{\dd}{\mathrm{d}}

\newcommand{\deriv}[3]{\frac{\partial\ifthenelse{\equal{#1}{}}{}{^{#1}}%
    #2}{\partial #3\ifthenelse{\equal{#1}{}}{}{^{#1}}}}
\newcommand{\dderiv}[3]{\frac{\dd\ifthenelse{\equal{#1}{}}{}{^{#1}}%
    #2}{\dd #3\ifthenelse{\equal{#1}{}}{}{^{#1}}}}

\newcommand{\msbar}{\ensuremath{\overline{\mbox{\abbrev{MS}}}}}
\newcommand{\drbar}{\ensuremath{\overline{\mbox{\abbrev{DR}}}}}

\newcommand{\tcalo}{\tilde{\calo}}

\newcommand{\calo}{\mathcal{O}}
\newcommand{\cale}{\mathcal{O}}
\newcommand{\PP}{\calo\calo}
\newcommand{\PE}{\calo E}
\newcommand{\EP}{E\calo}
\newcommand{\EE}{EE}

\newcommand{\myacrodef}[3]{\acrodef{#2}{#3}\newcommand{#1}{\ac{#2}}}

\acused{HKL}
\newcommand{\qcd}{\abbrev{QCD}} 
\acused{QCD} 

\allowdisplaybreaks

\usepackage{array}
\usepackage[capitalize]{cleveref}

\begin{document}

\preprint{TTK-21-58, TTP22-002, P3H-21-104}



\title{Effective electroweak Hamiltonian in the gradient-flow
  formalism}


\author{Robert V. Harlander}
\email{harlander@physik.rwth-aachen.de}
\affiliation{Institute for Theoretical
  Particle Physics and Cosmology,\\ RWTH Aachen University, 52056
  Aachen, Germany}
\author{Fabian~Lange}
\email{fabian.lange@kit.edu}
\affiliation{Institut f\"ur Theoretische
  Teilchenphysik, Karlsruhe Institute of Technology (KIT),
  Wolfgang-Gaede-Stra\ss{}e 1, 76128 Karlsruhe, Germany}
\affiliation{Institut f\"ur
  Astroteilchenphysik, Karlsruhe Institute of Technology (KIT),
  Hermann-von-Helmholtz-Platz 1, 76344 Eggenstein-Leopoldshafen,
  Germany}


\date{\today}

\begin{abstract}
  The effective electroweak Hamiltonian in the gradient-flow formalism is constructed for the current-current operators through next-to-next-to-leading order \qcd. The results are presented for two common choices of the operator basis.
  This allows for a consistent matching of perturbatively evaluated Wilson coefficients and non-perturbative matrix elements evaluated by lattice simulations on the basis of the gradient-flow formalism.
\end{abstract}


\maketitle

\myacrodef{\sftx}{SFTX}{small-flow-time expansion}
\myacrodef{\cmm}{CMM}{Chetyrkin-Misiak-M\"unz}
\myacrodef{\wrt}{w.r.t.}{with respect to}
\myacrodef{\vpf}{VPF}{vacuum polarization function}
\myacrodef{\vev}{VEV}{vacuum expectation value}
\myacrodef{\rg}{RG}{renormalization group}
\myacrodef{\gff}{GFF}{gradient-flow formalism}
\myacrodef{\ope}{OPE}{Operator Product Expansion}
\myacrodef{\ckm}{CKM}{Cabbibo-Kobayashi-Maskawa}
\myacrodef{\lhc}{LHC}{Large Hadron Collider}
\myacrodef{\uv}{UV}{ultraviolet} \myacrodef{\lo}{LO}{leading order}
\myacrodef{\nlo}{NLO}{next-to-leading order}
\myacrodef{\nnlo}{NNLO}{next-to-next-to-leading order}
\myacrodef{\llog}{LL}{leading logarithmic}
\myacrodef{\nll}{NLL}{next-to-leading logarithmic}
\myacrodef{\nnll}{NNLL}{next-to-next-to-leading logarithmic}
\myacrodef{\pdf}{PDF}{parton density function}
\myacrodef{\sm}{SM}{Standard Model}
\myacrodef{\bsm}{BSM}{beyond-the-\ac{SM}}
\myacrodef{\mssm}{MSSM}{Minimal Supersymmetric \ac{SM}}
\myacrodef{\susy}{SUSY}{Supersymmetry}
\myacrodef{\dreg}{DREG}{Dimensional Regularization}
\myacrodef{\dred}{DRED}{Dimensional Reduction}
\myacrodef{\emt}{EMT}{energy-momentum tensor}



\section{Introduction}

The \gff~\cite{Luscher:2010iy} offers a promising solution to the matching of
perturbative and non-perturbative calculations. A potential application is
flavor physics, where non-perturbative matrix elements are typically evaluated
using lattice regularization, while the Wilson coefficients are calculated
perturbatively in dimensional regularization. The idea is to express the
regular higher-dimensional operators of the effective electroweak Hamiltonian
in terms of \uv-finite flowed operators. The matching between the regular and
the flowed operators is perturbative and can be absorbed into flow-time
dependent Wilson coefficients. The application of this approach to the
energy-momentum tensor
through \nnlo{} \qcd{}~\cite{Suzuki:2013gza,Makino:2014taa,Harlander:2018zpi}
has already shown to give competitive results, see
e.g.\ \citeres{Iritani:2018idk,Taniguchi:2020mgg,Shirogane:2020muc}. More
recently, the matching matrix has also been calculated for the quark dipole
operators at \nlo\ \qcd~\cite{Rizik:2020naq,Mereghetti:2021nkt}, and for the hadronic vacuum
polarization through \nnlo\ \qcd~\cite{Harlander:2020duo}.

In \citere{Suzuki:2020zue} the matching matrix for the current-current
operators of the effective electroweak Hamiltonian has been calculated
at \nlo\ \qcd\ in the \drbar\ scheme. Here we present the \nnlo\ expression
for this quantity in the basis defined in \citere{Chetyrkin:1997gb} which
allows us to adopt the \msbar\ scheme with a fully anti-commuting
$\gamma_5$. We also provide the results for the non-mixing basis though. The
perturbative input for a consistent first-principles calculation of $K$- or
$B$-mixing parameters on the basis of the \gff\ is thus available. Once the
corresponding lattice input exists, it will be interesting to see how
the \gff\ approach applied to flavor physics compares to results obtained with
conventional approaches (see \citere{Aoki:2021kgd} for an overview).


\section{Operator basis}

The effective electroweak Hamiltonian can be written schematically as
\begin{equation}
  \label{eq:hamil} \mathcal{H}_\mathrm{eff} =
  - \frac{4G_\mathrm{F}}{\sqrt{2}} V_\mathrm{CKM} \, \sum_n
  C_n \calo_n
\end{equation}
where $G_\mathrm{F}$ denotes the Fermi constant, $V_\mathrm{CKM}$ comprises
the relevant elements of the \ckm\ matrix, and $C_n$ are the Wilson coefficients.
In this work we focus on the current-current operators and choose
\begin{equation}
  \label{eq:phys} \begin{split}
  \calo_1 &=
  - \left(\bar\psi_{1} \gamma_\mu^{\mathrm{L}}
  T^a \psi_{2}\right) \left(\bar\psi_{3} \gamma_\mu^{\mathrm{L}}
  T^a \psi_{4}\right) ,\\ \calo_2
  &= \left(\bar\psi_{1} \gamma_\mu^{\mathrm{L}} \psi_{2}\right)
  \left(\bar\psi_{3} \gamma_\mu^{\mathrm{L}} \psi_{4}\right)
  \end{split}
\end{equation}
as our operator basis~\cite{Chetyrkin:1997gb}, where we adopt the Euclidean metric and use the short-hand notation
\begin{equation}\label{eq:basis:hunt}
  \begin{split}
    \gamma_\mu^{\mathrm{L}}= \gamma_\mu\frac{1-\gamma_5}{2}\,.
  \end{split}
\end{equation}
Our convention for the color generators is
\begin{equation}\label{eq:easy}
  \begin{split}
    [T^a,T^b] = f^{abc}T^c\,,\quad \text{Tr}(T^aT^b) = -\tr\delta^{ab}\,,
  \end{split}
\end{equation}
with $f^{abc}$ real and totally anti-symmetric. Working in dimensional regularization with $D=4-2\ep$, loop corrections lead to contributions which are not proportional to the operators of
\cref{eq:phys}. They have to be attributed to
so-called evanescent operators which vanish for $D = 4$, but mix with
the physical operators at higher orders in perturbation
theory~\cite{Buras:1989xd}. Following \citere{Chetyrkin:1997gb}, we
choose
\begin{equation}
  \label{eq:basis:feed} \begin{split}
    \cale_1^{(1)} &=
  - \left(\bar\psi_{1} \gamma_{\mu\nu\rho}^{\mathrm{L}}
  T^a \psi_{2}\right) \left(\bar\psi_{3} \gamma_{\mu\nu\rho}^{\mathrm{L}}
  T^a \psi_{4}\right) - 16 \calo_1 ,\\ \cale_2^{(1)}
  &= \left(\bar\psi_{1} \gamma_{\mu\nu\rho}^{\mathrm{L}}
  \psi_{2}\right) \left(\bar\psi_{3}
  \gamma_{\mu\nu\rho}^{\mathrm{L}} \psi_{4}\right)
  - 16 \calo_2\,,\\
  \cale_1^{(2)} &=
  - \left(\bar\psi_{1} \gamma_{\mu\nu\rho\sigma\tau}^{\mathrm{L}}
  T^a \psi_{2}\right) \left(\bar\psi_{3}
  \gamma_{\mu\nu\rho\sigma\tau}^{\mathrm{L}}
  T^a \psi_{4}\right)\\&\quad - 20 \calo_1^{(1)} - 256 \calo_1 ,\\
  \cale_2^{(2)}
  &= \left(\bar\psi_{1} \gamma_{\mu\nu\rho\sigma\tau}^{\mathrm{L}}
  \psi_{2}\right) \left(\bar\psi_{3} \gamma_{\mu\nu\rho\sigma\tau}^{\mathrm{L}}
  \psi_{4}\right)
  \\&\quad- 20 \calo_2^{(1)} - 256 \calo_2 .
  \end{split}
\end{equation}
as evanescent operators, where $\gamma_{\rho\mu_1 \cdots
  \mu_n}^{\mathrm{L}} \equiv \gamma_\rho^{\mathrm{L}}\gamma_{\mu_1}
\cdots \gamma_{\mu_n}$. We will refer to the basis defined by
\cref{eq:phys,eq:basis:feed} as the \cmm-basis in what follows.


\section{Flowed operators}

In the \gff, one defines flowed gluon and quark fields
$B^a_\mu=B^a_\mu(t)$ and $\chi=\chi(t)$ as solutions of the
flow equations~\cite{Luscher:2010iy,Luscher:2013cpa}
\begin{equation}
  \begin{split}
    \partial_t B^a_\mu &= \mathcal{D}^{ab}_\nu G^b_{\nu\mu} + \kappa
    \mathcal{D}^{ab}_\mu \partial_\nu B^b_\nu\,,\\ \partial_t \chi
    &= \Delta \chi - \kappa \partial_\mu B^a_\mu T^a \chi\,,\\ \partial_t
    \bar \chi &= \bar \chi \overleftarrow \Delta + \kappa \bar
    \chi \partial_\mu B^a_\mu T^a\,,
    \label{eq:flow}
  \end{split}
\end{equation}
with the initial conditions
\begin{equation}
  \begin{split}
    B^a_\mu (t=0) = A^a_\mu\,,\qquad \chi (t=0)= \psi\,,
    \label{eq:bound}
  \end{split}
\end{equation}
where $A^a_\mu$ and $\psi$ are the regular gluon and quark fields,
respectively, and
\begin{equation}\label{eq:dleftright}
  \begin{split}
    \mathcal{D}^{ab}_\mu &= \delta^{ab}\partial_\mu - f^{abc}
    B_\mu^c\,,\qquad \Delta = (\partial_\mu + B^a_\mu T^a)^2\,,\\
    G_{\mu\nu}^a &= \partial_\mu B_\nu^a -
    \partial_\nu B_\mu^a + f^{abc}B_\mu^bB_\nu^c\,.
  \end{split}
\end{equation}
The parameter $\kappa$ is arbitrary and drops out of physical
quantities; we will set $\kappa=1$ in our calculation, because this
choice reduces the size of the intermediate algebraic expressions.

Our practical implementation of the \gff\ in perturbation theory follows
the strategy developed in \citere{Luscher:2011bx} and further detailed
in \citere{Artz:2019bpr}. On the one hand, it amounts to generalizing
the regular \qcd\ Feynman rules by adding flow-time dependent
exponentials to the propagators. The flow equations, \cref{eq:flow}, are
taken into account with the help of Lagrange multiplier fields which are
represented by so-called ``flow lines'' in the Feynman diagrams. They
couple to the (flowed) quark and gluon fields at ``flowed vertices'', which
involve integrations over flow-time parameters.

While the flowed gluon field $B^a_\mu$ does not require renormalization~\cite{Luscher:2010iy,Luscher:2011bx}, the flowed quark fields $\chi$ have to be renormalized~\cite{Luscher:2013cpa}.
The non-minimal renormalization constant $\Zchiring$ for the flowed quark
fields $\chi$ is defined by the all-order condition~\cite{Makino:2014taa}
\begin{equation}\label{eq:zchidef}
  \begin{split}
    \Zchiring\langle \bar\chi\overleftrightarrow{\mathcal{\slashed{D}}}
    \chi\rangle_0\bigg|_{m=0} &\equiv -\frac{2\nc}{(4\pi t)^2}\,,\\[.5em]
    \overleftrightarrow{\mathcal{D}}_\mu = \partial_\mu
    -&\overleftarrow{\partial}\!_\mu + 2B_\mu^a T^a\,,
  \end{split}
\end{equation}
where $\langle\cdot\rangle_0$ denotes the \vev.  The \nnlo{} result for
$\Zchiring$ can be found in \citere{Artz:2019bpr}.

The flowed operators are then defined by replacing the spinors $\psi_i$ by
renormalized flowed spinors $\Zchiring^{1/2}\chi_i$ in the regular
operators, i.e.
\begin{equation}\label{eq:flowed:craw}
  \begin{split}
      \tcalo_1 &=
  - \Zchiring^2\left(\bar\chi_{1} \gamma_\mu^{\mathrm{L}}
  T^a \chi_{2}\right) \left(\bar\chi_{3} \gamma_\mu^{\mathrm{L}}
  T^a \chi_{4}\right) ,\\
  \tcalo_2
  &= \Zchiring^2\left(\bar\chi_{1} \gamma_\mu^{\mathrm{L}} \chi_{2}\right)
  \left(\bar\chi_{3} \gamma_\mu^{\mathrm{L}} \chi_{4}\right)\,,
  \end{split}
\end{equation}
and analogously for the evanescent operators. Due to the damping character of
the flow time $t>0$, matrix elements of the flowed operators are \uv\ finite
after renormalization of the strong coupling and the quark masses. One can
thus treat them in four space-time dimensions, which also means that flowed
evanescent operators can be neglected. However, we prefer to keep them in our
formalism, because it makes the equations more symmetric. Furthermore, the
fact that they have to vanish provides a welcome consistency check on our
results.
The regular
evanescent operators are still needed in our calculation, which will be
described below.


\section{Small-Flow-Time Expansion}

In the limit $t\to 0$, the flowed operators behave as~\cite{Luscher:2011bx}
\begin{equation}\label{eq:esau}
  \begin{split}
      \begin{pmatrix}
        \tcalo(t) \\
        \tilde{E}(t)
  \end{pmatrix}
      \asymp \zeta^\bare(t) \begin{pmatrix} \calo \\ E
      \end{pmatrix}\,,
  \end{split}
\end{equation}
where we use the notation
\begin{equation}\label{eq:bohr}
  \begin{split}
    \calo &= (\calo_1,\calo_2)^\mathrm{T}\equiv
    (\calo_1^{(0)},\calo^{(0)}_2)^\mathrm{T}\,,\\
    E &= (\cale^{(1)}_1,\cale^{(1)}_2,\cale^{(2)}_1,\cale^{(2)}_2)^\mathrm{T}\,,
  \end{split}
\end{equation}
and analogously for the flowed operators. Here and in what follows, the
superscript ``$\bare$'' marks a ``bare'' quantity which will undergo
renormalization.  The symbol $\asymp$ is used to indicate that terms of $O(t)$ are neglected. It will be convenient to adopt the block-notation of \cref{eq:esau} also for matrices. For example, for the renormalized matching matrix we write
\begin{equation}\label{eq:gong}
  \begin{split}
    \zeta(t) =
    \left(
    \begin{matrix}
      \zeta_{\PP}(t) & \zeta_{\PE}(t)\\
      \zeta_{\EP}(t) & \zeta_{\EE}(t)
    \end{matrix}
    \right) ,
  \end{split}
\end{equation}
where the $2\times2$-submatrix $\zeta_{\PP}$ concerns only the physical
operators.

Since matrix elements of the bare operators
are divergent while those of flowed operators are finite,
the bare matching matrix $\zeta^\bare(t)$ is divergent as $D\to
4$. However, one may define renormalized operators whose matrix elements are finite:
\begin{equation}
  \label{eq:EW:ren}
  \begin{pmatrix}
    \calo \\
     E
  \end{pmatrix}^\ren
  = Z
  \begin{pmatrix}
    \calo \\
     E
  \end{pmatrix}
  \equiv
  \begin{pmatrix}
    Z_{\PP} & Z_{\PE} \\
    Z_{\EP} & Z_{\EE}
  \end{pmatrix}
  \begin{pmatrix}
    \calo \\
    E
  \end{pmatrix}
  ,
\end{equation}
where $Z$ is the corresponding renormalization matrix. It is common
to define all its entries in the $\msbar$ scheme, except for the
submatrix $Z_{\EP}$, whose finite part is chosen such that physical
matrix elements $\langle\cdot\rangle$ of evanescent operators vanish to all orders in
perturbation theory~\cite{Buras:1989xd,Dugan:1990df,Herrlich:1994kh}:
\begin{equation}\label{eq:EW:amoy}
  \begin{split}
    \langle E^\ren\rangle &=
    Z_{\EP}\langle \calo\rangle
    + Z_{\EE}\langle E\rangle \overset{!}{=}O(\ep)\,.
  \end{split}
\end{equation}
Inserting \cref{eq:EW:ren} into \cref{eq:esau}, it follows
that
\begin{equation}
  \label{eq:EW_SMT_ren}
  \zeta(t) = \zeta^\bare(t)Z^{-1} =
  \left(
  \begin{matrix}
    \zeta_{\PP}(t) & \zeta_{\PE}(t)\\
    \zeta_{\EP}(t) & \zeta_{\EE}(t)
  \end{matrix}
  \right)
\end{equation}
is finite at $D=4$. Since $\langle \tilde{E}(t)\rangle = O(\ep)$, the
renormalization condition in \cref{eq:EW:amoy} is equivalent to
\begin{equation}\label{eq:junr}
  \begin{split}
    \zeta_{\EP}(t)=O(\ep)\,.
  \end{split}
\end{equation}


\section{Calculation of the matching matrix}

For the calculation of the matching matrix $\zeta(t)$ we use the method of
projectors~\cite{Gorishnii:1983su,Gorishnii:1986gn}. This means that we
define a set of matrix elements
\begin{equation}\label{eq:calculation:cion}
  \begin{split}
    P^{(i)}_j[X] = \langle0|X|i,j\rangle\bigg|_{p=m=0}\,,
  \end{split}
\end{equation}
with $i\in\{0,1,2\}$ and $j\in\{1,2\}$, such that
\begin{equation}\label{eq:deli}
  \begin{split}
    P^{(i)}_j[\calo^{(i')}_{j'}] = \delta_{ii'}\delta_{jj'}\,,
    \end{split}
\end{equation}
where we remind the reader of the unified notation for physical
and evanescent operators defined in \cref{eq:bohr}.  In general, the
projectors could also involve derivatives w.r.t.\ masses and/or external
momenta, but this is not the case for the set of operators considered
here. Since all external mass scales are set to zero in
\cref{eq:calculation:cion}, it is sufficient to satisfy \cref{eq:deli}
at tree-level, because all higher perturbative orders on the
l.h.s.\ vanish in dimensional regularization.

The external states $|i,j\rangle$ are understood to project onto
left-handed spinors only.
Adopting an anti-commuting $\gamma_5$ thus eliminates all
$\gamma_5$'s from the traces at any order in the calculation~\cite{Chetyrkin:1997gb}.

\begin{figure}
  \begin{center}
    \begin{tabular}{ccc}
          \includegraphics[%
            width=.123\textwidth]%
                          {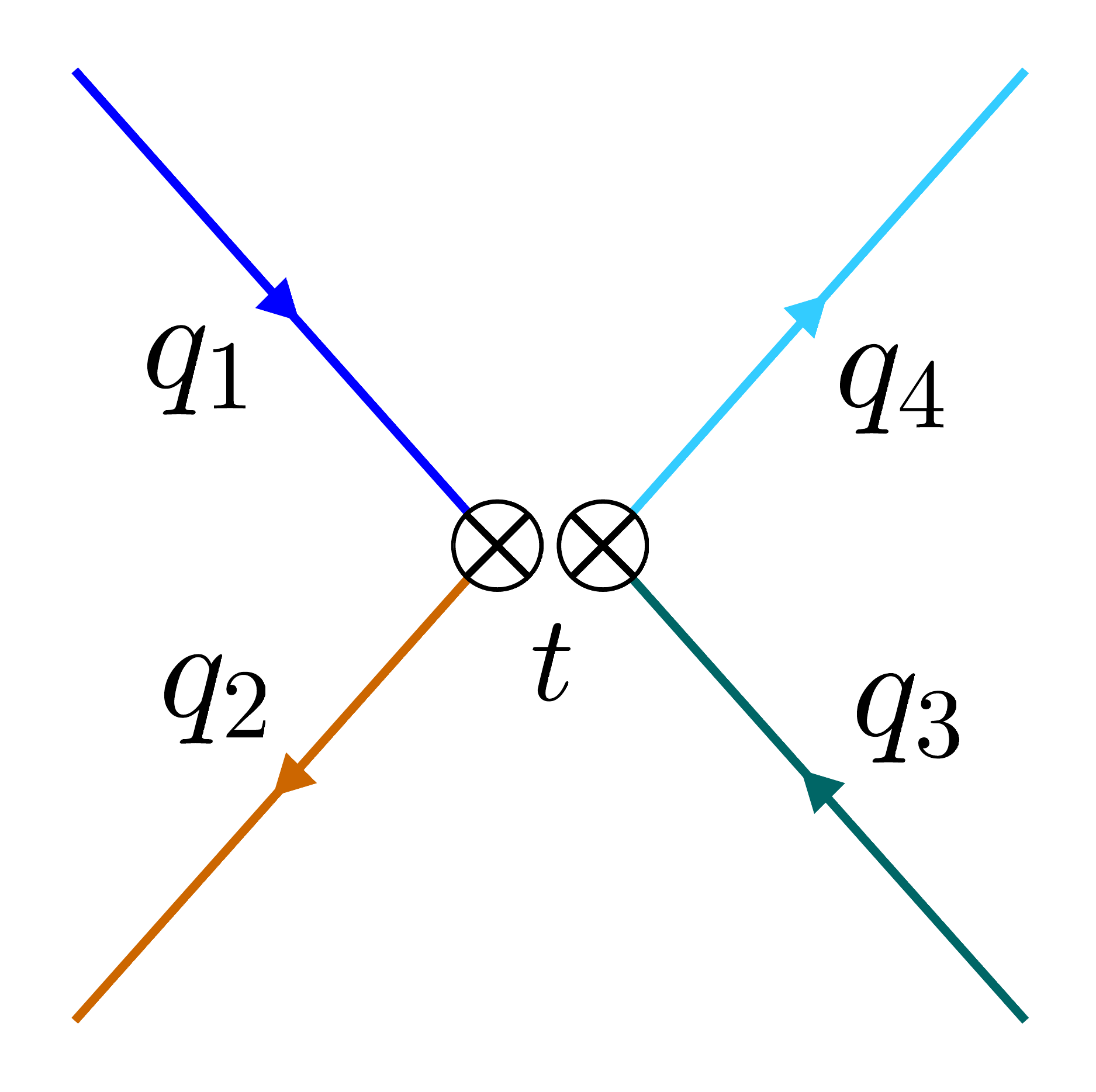} &
          \includegraphics[%
            width=.15\textwidth]%
                          {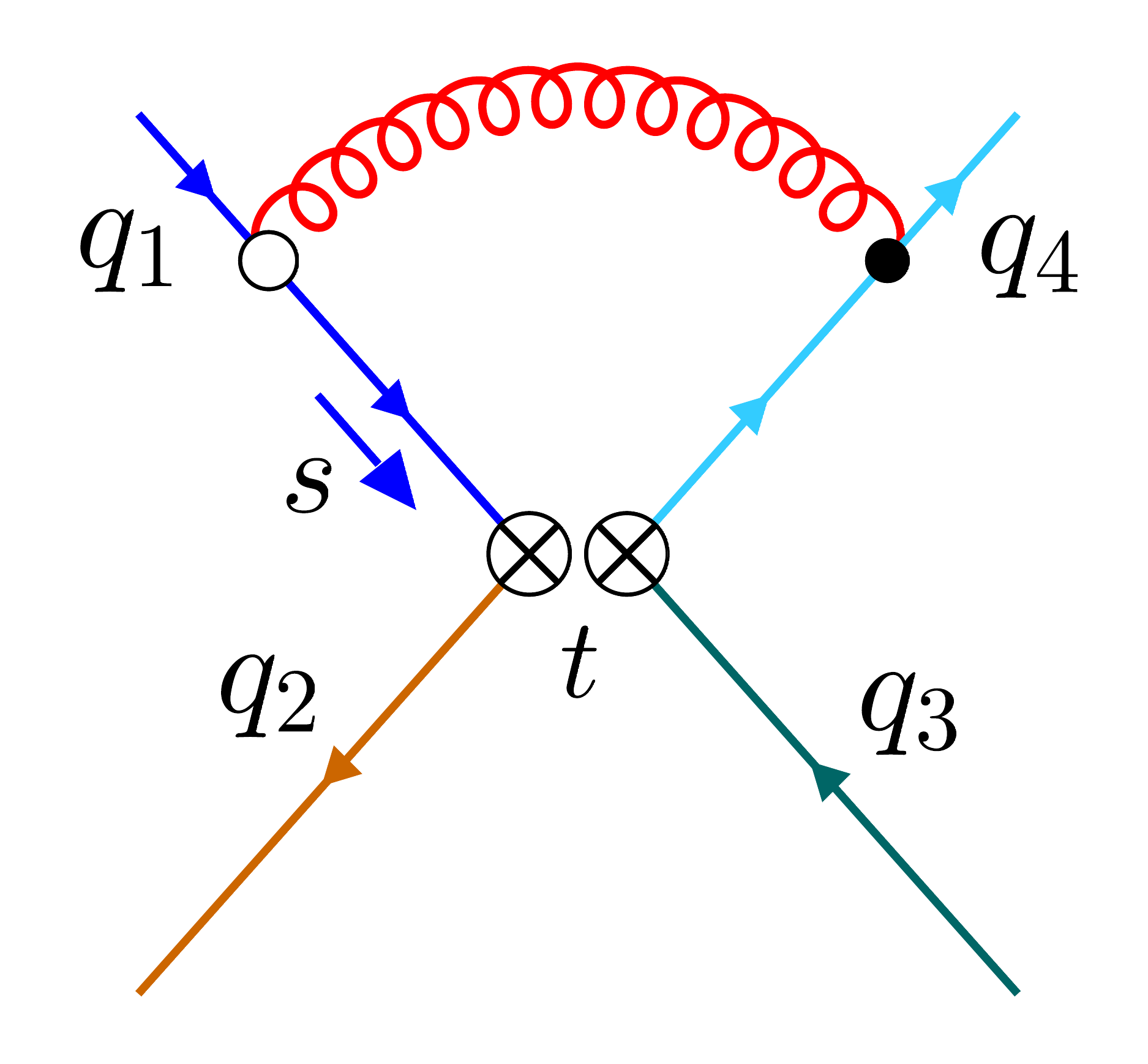} &
          \includegraphics[%
            width=.15\textwidth]%
                          {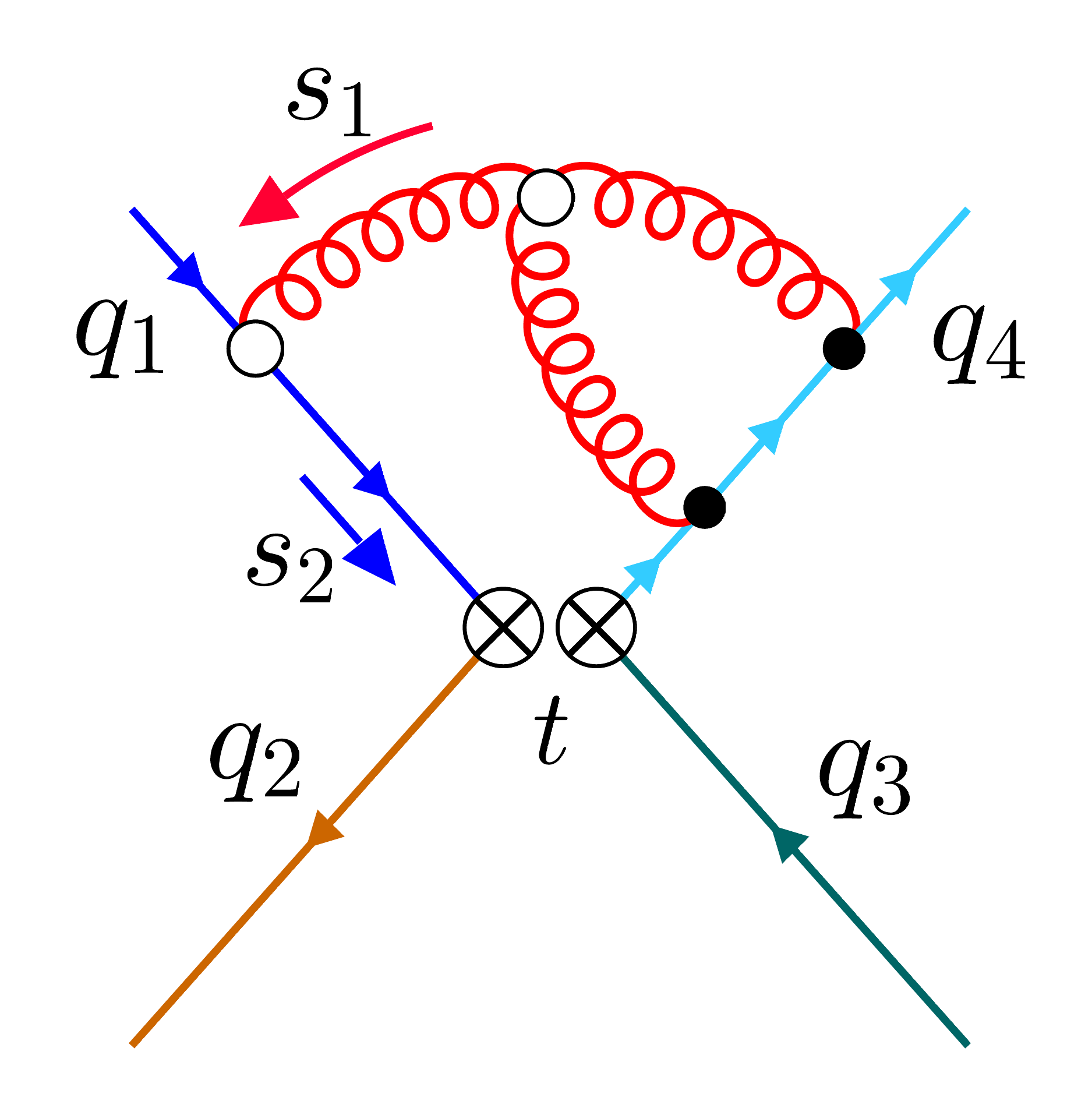}\\[-1em]
                            (a) & (b) & (c)
    \end{tabular}
      \caption[]{\label{fig:dias}\sloppy Sample diagrams contributing to the determination of the matching matrix $\zeta(t)$ at \lo, \nlo, and \nnlo\ \qcd. The circles denote ``flowed vertices'', lines with an arrow next to them denote ``flow lines'', and the label next to the arrow is a flow-time integration variable (see \citere{Artz:2019bpr} for details). The diagrams were produced with \texttt{FeynGame}~\cite{Harlander:2020cyh}.}
  \end{center}
\end{figure}

The bare matching matrix is obtained by applying the projectors to
\cref{eq:esau}:
\begin{equation}\label{eq:dory}
  \begin{split}
    \zeta^{\bare,(ii')}_{jj'}(t) = P^{(i')}_{j'}[\tcalo^{(i)}_j(t)]\,,
  \end{split}
\end{equation}
where the index notation should be self-explanatory.\footnote{For the
sake of clarity, let us point out that $\zeta^{(00)}_{jj'} \equiv
(\zeta_{\PP})_{jj'}$.}  Due to the fact that we restrict ourselves
to the case where all four quark flavors in the operator are different,
the Feynman diagrams contributing to the r.h.s.\ of this equation are
obtained by dressing the generic tree-level diagram in \cref{fig:dias}\,(a) by
virtual gluons and closed quark loops. Sample diagrams are shown in
\cref{fig:dias}\,(b) and (c).


%

%


For the actual evaluation of the diagrams, we adopt the setup based on
\texttt{q2e/exp}~\cite{Harlander:1998cmq,Seidensticker:1999bb}
described in \citere{Artz:2019bpr}. Specifically, we generate the
Feynman diagrams with
\texttt{qgraf}~\cite{Nogueira:1991ex,Nogueira:2006pq}, apply the
projectors, perform the traces, and simplify the algebraic expressions
within \texttt{FORM}~\cite{Vermaseren:2000nd,Kuipers:2012rf,vanRitbergen:1998pn}, and
reduce the resulting Feynman integrals to master integrals with the help
of \texttt{Kira+FireFly}~\cite{Maierhofer:2017gsa,Klappert:2020nbg,
  Klappert:2019emp,Klappert:2020aqs}. The master integrals are the same
as those found in \citere{Harlander:2018zpi}.


\section{Results}

\paragraph{\cmm\ basis.}
Performing the calculation and renormalization as described in the
previous sections, we find for the physical components of the
renormalized matching matrix through \nnlo{} in \qcd:
\begin{widetext}
\begin{equation}\label{eq:EW:glob}
  \begin{split}
    (\zeta^{-1})_{11}(t) &= 1
    +\api\, \left(4.212+ \frac{1}{2} \lmut\right)
    + \api^2 \bigg[
    22.72-0.7218\, \nf
    +\lmut \left(16.45-0.7576\, \nf\right)
    + \lmut^2 \left(\frac{17}{16}-\frac{1}{24}\, \nf\right)
    \bigg]\,,\\
    (\zeta^{-1})_{12}(t) &=
    \api\,\left(
    -\frac{5}{6}
    - \frac{1}{3}\lmut
    \right)
    + \api^2\,\bigg[
    -4.531 + 0.1576\,\nf
    + \lmut\,\left(-3.133 + \frac{5}{54}\,\nf\right)
    + \lmut^2\,\left(-\frac{13}{24} + \frac{1}{36}\nf\right)
    \bigg]\,,
     \\
     (\zeta^{-1})_{21}(t) &=
     \api\,\left(-\frac{15}{4} - \frac{3}{2}\,\lmut\right)
     + \api^2\,\bigg[-23.20
     + 0.7091\,\nf
     + \lmut\,\left(-15.22
     + \frac{5}{12}\,\nf\right)
     + \lmut^2\,\left(-\frac{39}{16} + \frac{1}{8}\,\nf\right)
     \bigg]\,,\\
     (\zeta^{-1})_{22}(t) &=
     1
     + 3.712\,\api
     + \api^2\,\bigg[
     19.47
     - 0.4334\,\nf
     + \lmut\,\left(11.75 - 0.6187\,\nf\right)
     + \frac{1}{4}\,\lmut^2
     \bigg]\,,
  \end{split}
\end{equation}
\end{widetext}
with $\api=\alpha_s(\mu)/\pi$ and $\lmut=\ln 2\mu^2t + \EulerGamma$, where $\alpha_s$ is the strong coupling renormalized in the \msbar\ scheme with $\nf$ quark flavors, $\mu$ the renormalization scale, and $\EulerGamma=0.577\ldots$ Euler's constant. For the sake of compactness, we set $\nc=3$ and $\tr=\tfrac{1}{2}$, and replaced transcendental coefficients by floating-point numbers. Analytical coefficients for a general SU($\nc$) gauge group are included in an ancillary file accompanying this paper.

Several observations support the correctness of this result. First of all, the literature expression for the renormalization matrix $Z$ defined through \cref{eq:EW:ren,eq:EW:amoy}~\cite{Chetyrkin:1997gb,
Gambino:2003zm,Gorbahn:2004my} not only
eliminates all \uv\ divergences from the matching matrix, but also nullifies its $E\calo$ component, see \cref{eq:junr}. Furthermore, we performed the calculation in $R_\xi$ gauge and found the result to be independent of the gauge parameter $\xi$. Yet another check concerns the switch to a different basis as described in the following.

\paragraph{Non-mixing basis.} It may be useful in physical applications to transform our result into the so-called non-mixing basis, defined such that the anomalous dimension matrix for the operators is diagonal.
The physical operators in that basis read
\begin{equation}\label{eq:opm}
  \calo_\pm = \frac{1}{2}\big[\big(\bar\psi_{1}^\alpha \gamma_\mu^\mathrm{L} \psi_{2}^\alpha\big) \big(\bar\psi_{3}^\beta \gamma_\mu^\mathrm{L} \psi_{4}^\beta\big) \pm \big(\bar\psi_{1}^\alpha \gamma_\mu^\mathrm{L} \psi_{2}^\beta\big) \big(\bar\psi_{3}^\beta \gamma_\mu^\mathrm{L} \psi_{4}^\alpha\big)\big]
\end{equation}
with the color indices $\alpha$, $\beta$.
The definition of the evanescent operators as well as the transformation matrices w.r.t.\ the \cmm\ basis are provided in \citere{Buras:2006gb} through \nnlo.\footnote{Note that the entry $\tfrac{8032}{75}$ in the matrix $\hat{V}$ in Eq.~(B.5) of \citere{Buras:2006gb} (Eq.~(A.8) in the arXiv version) should read $\tfrac{8032}{25}$.} We can easily evaluate the results in that basis by applying the corresponding transformation to the bare results for the projections obtained through \cref{eq:calculation:cion} and then performing the renormalization in complete analogy to the calculation for the \cmm\ basis. Alternatively, the transformation can be done at the level of the renormalized results by taking into account the required finite renormalization given in \citere{Buras:2006gb} to restore the renormalization scheme in the new operator basis~\cite{Chetyrkin:1997gb}. The fact that both ways lead to the same result and that the physical matching matrix $\zeta(t)$ between the \msbar\ renormalized and the flowed operators turns out to be diagonal in this basis is another strong check on our results. We find
\begin{widetext}
\begin{equation}\label{eq:EW:holm}
\begin{split}
  \zeta^{-1}_{++} (t) &=
  1
  +\api\left(
  2.796
  -\frac{1}{2} \lmut
  \right)
  +\api^2 \bigg[
  14.15
  -0.1739\, \nf
  + \lmut \left(
  6.509
  -0.4798\, \nf
  \right)
  +\lmut^2 \left(
  -\frac{9}{16}
  +\frac{1}{24}\, \nf
  \right)
  \bigg]\,,\\
  \zeta^{-1}_{--} (t) &=
  1+\api
  \left(5.546+\lmut\right)
  +\api^2
  \bigg[
  32.01
  -0.9524 \,\nf
  +\lmut
  (21.23\, -0.8965 \,\nf)
  +\lmut^2 \left(\frac{15}{8} - \frac{1}{12} \,\nf\right)
  \bigg]\,,
\end{split}
\end{equation}
\end{widetext}
where the same notation as in \cref{eq:EW:glob} is adopted.\footnote{An immediate comparison of this result to the \nlo\ expression of \citere{Suzuki:2020zue} is not possible, because the latter is obtained in the \drbar\ scheme.} Again, analytical results are provided in the ancillary file.\footnote{%
  Since the non-mixing basis in \citere{Buras:2006gb} was constructed for $\nc = 3$, we also insert this value for $\zeta^{-1}_{++}$ and $\zeta^{-1}_{--}$ in the ancillary file, and in addition set $\tr=\tfrac{1}{2}$. A non-mixing basis for general $\nc$ could be easily constructed from our results though.}

We note in passing that the matching matrix also determines the small-$t$ behavior of the flowed operators through the equation~\cite{Harlander:2020duo}
\begin{equation}\label{eq::heed}
\begin{split}
t\partial_t\tilde{\mathcal{O}}(t) = \tilde{\gamma}(t)\tilde{\mathcal{O}}(t)\,,\qquad
\tilde{\gamma}(t) = (t\partial_t\zeta(t))\zeta^{-1}(t)\,.
\end{split}
\end{equation}
These equations hold in any basis, of course.


\section{The effective Hamiltonian in the gradient-flow formalism}

Inverting the small-flow-time expansion in \cref{eq:esau}, one can write the
Hamiltonian as
\begin{equation}
  \label{eq:hamilflow} \mathcal{H}_\mathrm{eff} \asymp
  - \frac{4G_\mathrm{F}}{\sqrt{2}} V_\mathrm{CKM} \, \sum_n
  \tilde{C}_n(t) \tcalo_n(t)\,,
\end{equation}
where the flowed Wilson coefficients are given by
\begin{equation}\label{eq:arad}
  \begin{split}
    \tilde{C}_n(t) = \sum_{m}C_m^\ren\,\zeta^{-1}_{mn}(t)\,,
  \end{split}
\end{equation}
with $\zeta(t)\equiv \zeta_{\PP}(t)$ the physical part of the matching matrix,
and $C_n^\ren=\sum_{m}C_m(Z^{-1})_{mn}$ the renormalized regular Wilson
coefficients. It is important to evaluate $C^\ren$ and $\zeta^{-1}(t)$ in the
same renormalization scheme, including the treatment of $\gamma_5$ and the
choice of (regular) evanescent operators. The flowed
coefficients $\tilde{C}(t)$, on the other hand, are scheme and renormalization
scale independent (up to higher orders in perturbation theory).  Since also
the flowed operators $\tcalo(t)$ are scheme and renormalization scale
independent, \cref{eq:hamilflow} allows one to combine perturbatively
calculated Wilson coefficients with non-perturbative matrix elements without
scheme transformation.

In order to avoid large logarithms, after matching the
$C_m^\ren=C_{m}^\ren(M_W, \mu)$ to the
\sm\ at $\mu \sim M_W$, they should be evolved down to $\mu\sim\sqrt{1/t}$
using the standard renormalization group
equation~\cite{Buchalla:1995vs,Gorbahn:2004my}, where $t$ is sufficiently
large to warrant small uncertainties in the lattice
calculation. Alternatively, one may choose to perform the evolution to large $t$
at the level of the flowed coefficients, using
\begin{equation}
  t\partial_t\tilde{C}_m(t) = - \sum_n \tilde{C}_n(t)\tilde\gamma_{nm}(t)
\end{equation}
with $\tilde\gamma(t)$ defined
in \cref{eq::heed}. The compatibility of both approaches is left for future
investigation.

For $|\Delta F| = 1$ processes, the Wilson coefficients $C^\ren_m$ in
the \cmm\ basis for the \sm{} can be found
in \citeres{Bobeth:1999mk,Gorbahn:2004my} through \nnlo{}.  Thus, when
neglecting penguin contributions, re-expanding the r.h.s.\ of \cref{eq:arad}
through \nnlo\ using the results for $\zeta^{-1}(t)$ above, directly gives the
flowed Wilson coefficients to the same order.  For $|\Delta F| = 2$ processes,
the physical basis reduces to just one operator due to a Fierz identity.  In
this case, the \sm{} Wilson coefficient is known
through \nlo{}~\cite{Buchalla:1995vs}, with two contributions for kaon mixing
known through \nnlo{}~\cite{Brod:2010mj,Brod:2011ty}.


\section{Conclusions and outlook}

We calculated the matching matrix of the current-current operators in the electroweak effective Hamiltonian to their flowed counterparts through \nnlo\ \qcd. We presented the results in the \cmm\ and the non-mixing bases and performed a number of checks on their correctness. Our results can directly be applied to $K$- or $B$-meson mixing, for example. Their generalization, in particular the inclusion of penguin operators, is work in progress.
It remains to be seen how the \gff\ approach to flavor physics compares to conventional calculations.



\begin{acknowledgments}


We are indebted to Marvin Gerlach, Martin Lang, Antonio Rago, Andrea Shindler, and Benjamin Summ for valuable comments and discussions. Special thanks go to Martin L\"uscher and the CERN theory group for setting off this project, and to Ulrich Nierste for valuable clarifications, especially concerning evanescent operators. We also thank Martin L\"uscher for his comments on the manuscript.
This work was supported by \textit{Deutsche Forschungsgemeinschaft
  (DFG)} through project
\href{http://gepris.dfg.de/gepris/projekt/386986591}{HA 2990/9-1} and
grant \href{http://gepris.dfg.de/gepris/projekt/396021762?language=en}{396021762} -- \href{http://p3h.particle.kit.edu/start}{TRR 257} ``Particle Physics Phenomenology after the
Higgs Discovery''.

\end{acknowledgments}



\bibliography{paper}

\end{document}